\input{epsf}
\documentstyle[aps,preprint]{revtex}
\begin{document}
\tightenlines

\title{On the thresholds, probability densities, and critical exponents of Bak-Sneppen-like models}

\author{Guilherme J. M. Garcia$^*$ and Ronald Dickman$^\dagger$\\
 {\small Departamento de F\'\i sica, Instituto de Ci\^encias Exatas}\\
 {\small Universidade Federal de Minas Gerais, Caixa Postal 702}\\
  {\small CEP 30123-970, Belo Horizonte - Minas Gerais, Brazil} }

\date{\today}

\maketitle
\vskip 0.5truecm

\begin{abstract}

We report a simple method to accurately determine the threshold and the exponent
$\nu$ of the Bak-Sneppen model and also investigate the BS universality class. 
For the random-neighbor version of the BS model, we find the threshold
$x^*=0.33332(3)$, in agreement with the exact result $x^*=1/3$ given
by mean-field theory. For the one-dimensional original model, we find $x^*=0.6672(2)$
in good agreement with the results reported in the literature;
for the anisotropic BS model we obtain $x^*=0.7240(1)$.
We study the finite size effect $x^*(L)-x^*(L \to \infty) \propto L^{-\nu}$, 
observed in a system with $L$ sites, and find $\nu = 1.00(1)$ for the 
random-neighbor version, $\nu = 1.40(1)$ for the original model,
and $\nu=1.58(1)$ for the anisotropic case. 
Finally, we discuss the effect of defining the extremal site as the one 
which minimizes a general function $f(x)$, instead of simply $f(x)=x$ as in the
original updating rule. We emphasize that models with extremal dynamics have
singular stationary probability distributions $p(x)$.  Our simulations indicate the existence
of two symmetry-based universality classes.

\vspace{1em}

\noindent PACS: 05.65.+b, 02.30.Ks, 05.40.-a, 87.10.+e
\vspace{1em}

\noindent keywords: Bak-Sneppen model; threshold; finite-size scaling;
anisotropic BS model; critical exponents; universality;
\vspace{1em}

\noindent$^*$ Electronic address: gjmg@fisica.ufmg.br\\
$^\dagger$ Electronic address: dickman@fisica.ufmg.br\\

\noindent Corresponding author: Ronald Dickman\\
telephone: 55-31-3499-5665\\
FAX: 55-31-3499-5600

\end{abstract}

\newpage

\section{Introduction}

Power-law distributions are often found in natural and social systems.
Examples include the distributions of 
rain event sizes and durations \cite{Peters1,Peters2,Dickman1,lovejoy}, 
earthquake sizes \cite{terremotos,BTW}, 
internet connections \cite{pastor}, 
and food webs \cite{Pietronero}.
Why do such diverse phenomena exhibit scale invariance, as reflected in
power-law distributions?
A concept introduced by Bak and collaborators to account in part for
scale-invariance in nature is self-organized criticality (SOC), which
asserts that ``systems that are far from equilibrium evolve to a dynamical attractor poised
at criticality'' \cite{Paczuski}.

Over the last two decades, many models exhibing self-organized criticality have been
proposed:
sandpile models \cite{BTW,manna,manna2}, 
the OFC model for earthquakes \cite{OFC},
forest fire models \cite{forest fire}, 
the ballistic deposition model of rough surfaces \cite{ballistic},
and the Bak-Sneppen evolution model \cite{Bak and Sneppen:1993}.
Many SOC models are defined in such a manner that
they are trapped at criticality \cite{Dickman et al:2000}.
In particular, sandpile models are maintained at the critical point
by ``slow driving in a system exhibiting an absorbing-state phase transition 
with a conserved density'' \cite{Dickman et al:2000}.
A different path to SOC is so-called extremal dynamics. In this case, the system is critical
provided that an `extremal condition' is respected \cite{Dickman et al:2000,head,Garcia and Dickman}.

The Bak-Sneppen evolution model is one of the simplest models
exhibiting SOC via extremal dynamics. For this reason, it has atracted much attention
in the statistical physics community and has been studied through various approaches,
including simulation \cite{Grassberger:1995,Rios:1998,Boettcher:2000},
theoretical analysis \cite{Meester:2002,Li and Cai:2000,Dorogovtsev:2000},
probabilistic analysis (run time statistics) \cite{Caldarelli:2002,Felici:2001},
renormalization group \cite{Marsili:1994,Mikeska:1997},
field theory \cite{Paczuski:1994}
and mean-field theory \cite{Dickman et al:2000,head,Garcia and Dickman,Pismak:1997}.
Some variants have been proposed, for example the anisotropic BS model
\cite{Head:1998,Maslov:1998}.

In this paper, we investigate the scaling behavior of the Bak-Sneppen model.
We study the finite size scaling $x^*_L-x^*_\infty = k L^{-\nu}$, where $x^*_L$ is the threshold
for a one-dimensional system of $L$ sites and $k$ is a constant. This provides an accurate estimate of the
threshold $x^*_\infty$ and of the exponent $\nu$. We also discuss which
modifications of the updating rule preserve the universality class.
In agreement with Ref. \cite{Garcia and Dickman}, we find that extremal dynamics leads
to a singular stationary
distribution $p(x)$, and that models with symmetric updating rules belong 
to the same universality class of the original model. The specific form 
of the distribution $p(x)$ is, however, nonuniversal.

The balance of this paper is organized as follows. 
In Sec. II, we define the BS model along with the random-neighbor and the anisotropic BS versions.
We define also a ``cosine BS model''.
We calculate the threshold and the exponent $\nu$ of the first three models in Sec. III. 
In Sec. IV, we study the BS universality class. We summarize our conclusions in Sec. V.

\section{Models}

The Bak-Sneppen model \cite{Bak and Sneppen:1993} is defined on a d-dimensional lattice with 
$L^d$ sites and periodic boundaries. Each site bears a real-valued variable $x_i \in [0,1)$. 
At time zero, all the $x_i$ are assigned
independent random values drawn from a distribution uniform on
$[0,1)$. At each step of the evolution, the site $x_m$, that is the global minimum of 
the $\{ x_i \}$, along with its $2d$ nearest neighbors, are replaced by new random 
numbers taken from the same distribution. This process is repeated indefinitely.

In the random-neighbor version of the model, all sites are considered neighbors and,
at each time step, the site $m$ along with two randomly chosen neighbors are substituted
with new random numbers. All the other definitions remain the same.
A further variant of the model is the anisotropic BS model. In this case, 
only the site $m$ and its {\sl right} neighbor are replaced at each time step.
We also study a modified Bak-Sneppen model, that differs from the original in the
way the site $x_m$ is identified. We baptize this the ``cosine BS model''.
Instead of minimizing $f(x)=x$ as in the original model, 
$x_m$ minimizes $f(x)=\cos(2n\pi x)$ for some integer $n$, so that $f$
has $n$ minima on the interval [0,1].
Finally, in order to investigate the effect of symmetry in the updating rule,
we define an ``anisotropic cosine BS model'', 
in which we have the spatial anisotropy of the anisotropic BS model and
the extremal site defined as in the cosine BS model.

\section{Determining the threshold}

Simulation of the BS model shows that, after a transient time, the system achieves a
statistically stationary state in which the density of $x_i$ vanishes for $x$ less than a
certain threshold $x^*$ and is uniform above $x^*$. In other words, for an infinite system 
and in the stationary state, the distribution $p(x)$ is the step-fuction

\begin{equation}
p(x) =
\left\{ 
\begin{array}{c}
0 ~ , ~~~ x<x^* ~~ \\ 
C ~ , ~~~ x>x^* ~~.
\end{array}
\right.  
\end{equation}

This singular probability density is associated with scaling properties reminiscent of
a critical point; the upper critical dimension $d_c=4$ \cite{Boettcher:2000}.
The Bak-Sneppen model does not appear to belong to any of the known
universality classes of critical phenomena, and may be said to define
its proper class.

\begin{figure}
\epsfysize=5cm
\epsfxsize=6cm
\centerline{
\epsfbox{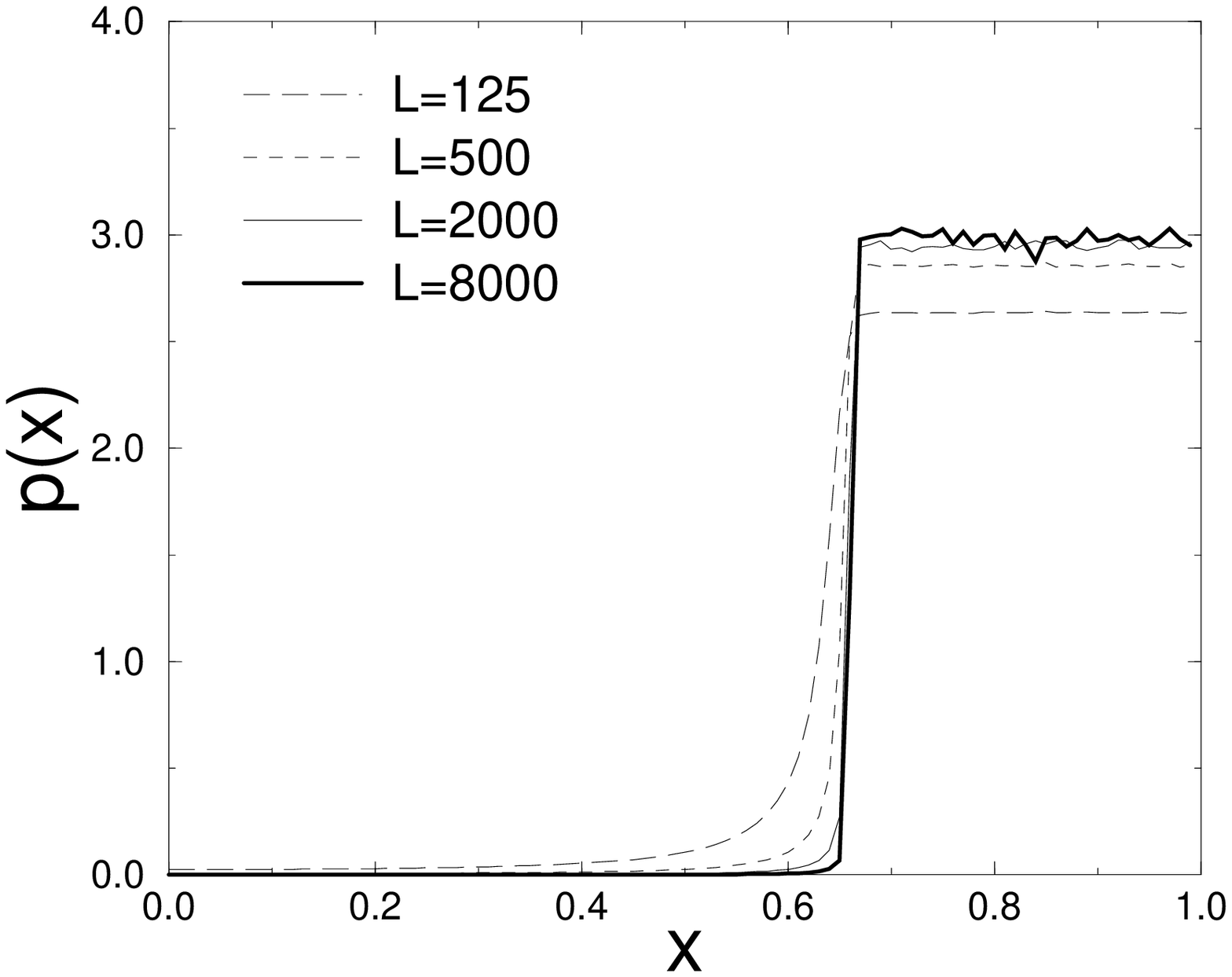}}
\caption{Stationary density $p(x)$ for the original Bak-Sneppen model in one dimension,
for various system sizes.}
\label{fig1}
\end{figure}

Accurate determinations of $x^*$ were reported by Grassberger \cite{Grassberger:1995}
and Paczuski and coworkers \cite{Paczuski}, who found $x^*=0.66702(8)$ and $x^*=0.66702(3)$,
respectively. We introduce here a third method to determine $x^*$. 
For a finite system (see Fig. \ref{fig1}), the distribution $p(x)$ is continuous,
i.e., the step function singularity is rounded \cite{note1}. 
Nevertheless, as is apparent from the figure, $p(x) = C_L$, constant, for $x$ well above the
transition region.   If $p(x)$ is a step function,
normalization implies that $C(1-x^*)=1$.
This suggests that we define a finite-system threshold via $x^*_L=1-1/C_L$. 
If we now assume that
the Bak-Sneppen model obeys the finite-size scaling relation

\begin{equation}
x^*_L-x^*_\infty = k L^{-1/\nu},
\label{eq2}
\end{equation} 

\noindent where $k$ is a constant and $x^*_\infty$ is the threshold for an infinite 
system, we have a simple method for determining the threshold of 
Bak-Sneppen-like models.

We simulate the Bak-Sneppen model, the anisotropic BS model, and the random-neighbor
version on lattices of $L=$ 63, 125, 250, 500, 1000, 2000, 4000 and 8000 sites
(see Figure \ref{fig1}). We estimate the probability density $p(x)$ on the
basis of a histogram of barrier frequencies, dividing [0,1] into 100 subintervals.
Histograms are accumulated after $N_{st}$ time steps, as required for the system 
to relax to the stationary state. (We find that  
for the original and the anisotropic models $N_{st}=10^8$
sufficient to reach the stationary states for even the largest system studied,
while for the random-neighbor version $N_{st}=10^7$ is sufficient.) 
We make the histogram every $L$ time steps in runs of $10^9$
steps total.

\begin{figure}
\epsfysize=5cm
\epsfxsize=6cm
\centerline{
\epsfbox{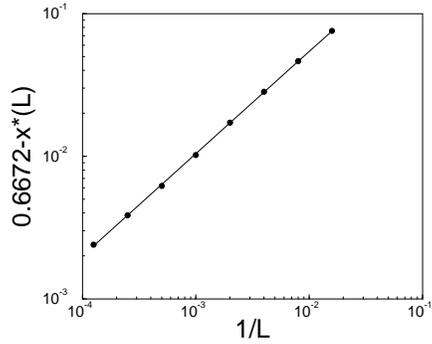}}
\caption{Power law fit for ($x^*_\infty-x^*_L$) vs. $1/L$ for the original BS model. The slope
of the straight line is 1.40.}
\label{fig2}
\end{figure}

To test the accuracy of the method described above, we applied it to the random-neighbor
version of the model, for which the exact result $x^*=1/3$ is known from mean-field theory 
\cite{Dickman et al:2000,head,Garcia and Dickman,Pismak:1997}.
Analyzing our data for $x^*_L$ {\sl vs.} $L$ using Eq. (\ref{eq2}), 
we find $x^*=0.33332(3)$ and $\nu=1.00(1)$. 
This results agrees with the mean-field theory prediction and validates our method.
For the original BS model, we find $x^*=0.6672(2)$ and $\nu = 1.40(1)$.
(The power law fit is shown in Fig. \ref{fig2}.)
In accord with this are the results $x^*=0.66702(8)$ reported by Grassberger \cite{Grassberger:1995},
$x^*=0.66702(3)$ found by Paczuski and collaborators \cite{Paczuski},
and $\nu=1.36$ \cite{Paczuski2}. Finally, for the anisotropic BS model, we obtain $x^*=0.7240(1)$
and $\nu=1.58(1)$. To our knowledge, this is the first result for the threshold and the
exponent $\nu$ of the anisotropic BS model.

\section{The BS Universality Class}

As noted in the Introduction, extremal dynamics is one path to a self-organized critical state. 
Two questions therefore arise: 1) What are the trademarks of extremal dynamics? and
2) What other models belong to 
the universality class of the Bak-Sneppen model? In order to answer these questions, 
Garcia and Dickman \cite{Garcia and Dickman} studied the effects of updating via $x_i' = x_i^\alpha$
(with $\alpha = 1/2$ or 2), instead of replacing these variables with random numbers. They found that
these updating rules lead to quite distinct stationary probability densities $p(x)$,
which nonetheless exhibit one or more discontinuities. 
Another caracteristic shared by all the variants studied in \cite{Garcia and Dickman}
is that the extremal site always
belongs to the ``prohibited region'' for which $p(x)=0$.
Although $p(x)$ and consequently the threshold are nonuniversal, all variants studied
belong to the universality class of the original model. Moreover, it has been shown 
\cite{Maslov:1998,Head and Rodgers:1998,Garcia and Dickman2}
that models that break the symmetry of the original model fall into the universality class of the
anisotropic BS model.

A related question is the robustness of BS scaling under modifications of extremal
dynamics.  It is known that scale invariance is lost when the extremal condition is relaxed,
for example by introducing a nonzero probability for choosing a nonextremal site to be
updated \cite{Dickman et al:2000,head,Garcia and Dickman}.  In this context it is 
interesting to note that updating the $M$ smallest barriers (with $M>1$), at each 
step changes the scaling properties drastically \cite{Datta}.

\begin{figure}
\epsfysize=5cm
\epsfxsize=6cm
\centerline{
\epsfbox{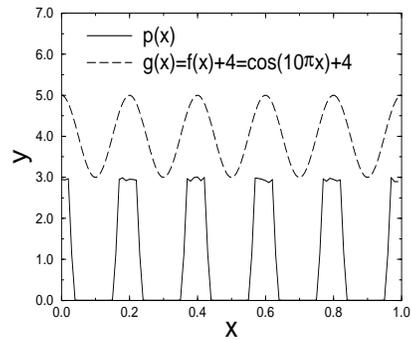}}
\caption{Solid line: stationary density $p(x)$ for the cosine BS model ($n=5$)
for a system of $L=2000$ sites. Dashed line: function $f(x)=\cos(10\pi x)$ 
that is minimized in the cosine BS model. ($f(x)$ is shifted for better 
visualization.) For further details, see Section II.}
\label{fig3}
\end{figure}

As representatives of a further broad class of variants of the  Bak-Sneppen model, 
we study here the isotropic and the anisotropic cosine BS models proposed in Sec. II.
We simulate these models (with $n=5$) on a lattice of $L=2000$ sites and find the 
distribution $p(x)$ shown in figure \ref{fig3} for the isotropic case.
(In the anisotropic version, a similar distribution is found.)
It is natural to regard the rounding of the singularities as a finite-size
effect. Thus we find evidence that these updating rules also lead to singular probabilty
densities. We have also investigated the distribution $P_J(r)$ that sucessive updated
sites are separated by a distance $r$. This obeys the power law $P_J(r) \sim r^{-\pi}$
with $\pi=3.23(2) $ in the original model \cite{Head:1998}, and $\pi=2.401(2)$ in the
anisotropic BS model \cite{Maslov:1998}.  
We performed simulations of the cosine BS models using $10^8$ time steps, 
yielding $\pi=3.2(1)$ for the symmetric updating rule and
$\pi=2.38(3)$ for asymmetric updating rule.
Therefore a dynamics that respects the symmetry of the original model
(that is, spatial isotropy), and that does not introduce any new conserved quantity,
leads to the same exponents as the original model. On the other hand,
asymmetric dynamics leads to the anisotropic BS model universality class.

\section{Conclusions}

We verify the finite-size scaling relation $x^*_L-x^*_\infty = k L^{-\nu}$
in the Bak-Sneppen model, and use it
to obtain accurate values for the threshold and for the exponent $\nu$
of the original, anisotropic and random-neighbor versions.
We also study the effect of defining the extremal site as the one that
minimizes a function $f(x)$ with multiple minima. Simulations show that
this does not affect the universality class and strengthens the result
of Ref. \cite{Garcia and Dickman} that extremal dynamics leads to singular
probability densities. Finally, our results indicate that variants that 
preserve the spatial isotropy of the original model belong to the same 
universality class, while models which break that symmetry fall into
the anisotropic BS model universality class.

\bigskip
\noindent {\bf ACKNOWLEDGMENTS}
\smallskip

\noindent We thank CNPq and CAPES, Brazil, for finacial suport.

\newpage

\end{document}